\documentclass[journal]{IEEEtran}

\usepackage{graphicx}
\graphicspath{{fig/}}
\usepackage{cite}
\usepackage{amsmath,amssymb,amsfonts}
\usepackage{algorithmic}
\usepackage{graphicx}
\usepackage{textcomp}
\usepackage{subfigure}
\usepackage[ruled]{algorithm2e}
\usepackage{multirow}
\usepackage{tabularx}
\usepackage{bm}
\usepackage{stfloats}
\usepackage{diagbox}
\usepackage{slashbox}
\usepackage{flushend}
\usepackage{balance}
\usepackage{soul}
\usepackage{float}

\soulregister\cite7
\soulregister\citep7
\soulregister\citet7
\soulregister\ref7
\soulregister\pageref7

\usepackage{xcolor,soul,framed} 

\colorlet{shadecolor}{yellow}

\begin{document}
	\title{Joint Cluster Head Selection and Trajectory Planning in UAV-Aided IoT Networks by Reinforcement Learning with Sequential Model}	
	\author{Botao~Zhu,~\IEEEmembership{}
		Ebrahim~Bedeer,~\IEEEmembership{Member,~IEEE,}
		~Ha H. Nguyen,~\IEEEmembership{Senior Member,~IEEE,} Robert~Barton,~\IEEEmembership{Member,~IEEE,} and Jerome Henry,~\IEEEmembership{Senior Member,~IEEE}
		\thanks{B. Zhu, E. Bedeer, and H. H. Nguyen are with the Department of Electrical and Computer Engineering, University of Saskatchewan, Saskatoon, Canada S7N5A9. Emails: \{botao.zhu, e.bedeer, ha.nguyen\}@usask.ca.}
		\thanks{R. Barton and J. Henry are with Cisco Systems Inc. Emails: \{robbarto, jerhenry\}@cisco.com.}
 		\thanks{Copyright (c) 20xx IEEE. Personal use of this material is permitted. However, permission to use this material for any other purposes must be obtained from the IEEE by sending a request to pubs-permissions@ieee.org.}
	}

	\maketitle
	\begin{abstract}
			 Employing unmanned aerial vehicles (UAVs) has attracted growing interests and emerged as the state-of-the-art technology for data collection in Internet-of-Things (IoT) networks. In this paper, with the objective of minimizing the total energy consumption of the UAV-IoT system, we formulate the problem of jointly designing the UAV's trajectory and selecting cluster heads in the IoT network as a constrained combinatorial optimization problem which is classified as NP-hard, and challenging to solve. We propose a novel deep reinforcement learning (DRL) with a sequential model strategy that can effectively learn the policy represented by a sequence-to-sequence neural network for the UAV's trajectory design in an unsupervised manner. Through extensive simulations, the obtained results show that the proposed DRL method can find the UAV's trajectory that requires much less energy consumption when compared to other baseline algorithms and achieves close-to-optimal performance. In addition, simulation results show that the trained model by our proposed DRL algorithm has an excellent generalization ability to larger problem sizes without the need to retrain the model. 	
	\end{abstract}
	

	\section{Introduction}
	\label{sec:introduction}
	 \IEEEPARstart{T}{he} Internet-of-Things (IoT) is a system that connects a massive number of devices to the Internet, which is rapidly changing the way we live in almost every field \cite{S. Chen}. Wireless Sensor Networks (WSNs) are viewed as the basic component of IoT. WSN can integrate the physical world with the information world to expand the functions of existing networks and the ability of humans to understand the world. A typical WSN is composed of a large number of devices deployed over a geographical area for monitoring physical events. These devices form a multi-hop self-organizing network to monitor, sense, and collect the information of the target area, and transfer the collected data to users for processing \cite{H. Xie}. However, WSNs have gradually merged with different applications and appeared in various new forms, such as industrial IoT \cite{K. R. Choo}, Internet of vehicles (e.g., 5G long distance WSN) \cite{Q. Zhang}, smart home (e.g., short distance WSN) \cite{C. Tseng}, environmental monitoring (e.g., autonomous underwater vehicles network) \cite{D. Wei et al}, intelligent manufacturing system, etc.

	 Traditional long-distance multi-hop communication requires high energy consumption of end devices, and because of the limited energy resources of end devices in IoT networks, it leads to shortening the network lifetime. In \cite{Y. Kim}, an ant optimization based routing algorithm is proposed to dramatically reduce the energy consumption of networks. In \cite{D. Lin}, the authors consider economic theory and compressive sensing theory to propose an energy-saving routing algorithm to extend the lifetime of WSNs. These energy saving techniques only focus on designing the routing algorithms to reduce the energy consumption of networks. Recently, the use of unmanned aerial vehicles (UAVs) has received increasing attention due to their high flexibility and high maneuverability. Compared with the conventional IoT networks that use static multi-hop data collection methods, UAV-enabled IoT networks dispatch a UAV to collect data from ground IoT devices based on the planned UAV's trajectory \cite{M. Samir}, which can effectively reduce the energy consumption of devices. {However, any increase in the flight time or distance of the UAV for a given data collection mission will increase its energy consumption. Hence, there is a need to carefully design the UAV trajectory to minimize the overall energy consumption of the wireless network.}

	 There is rich literature concerning the problem of energy consumption in UAV-aided wireless networks. By jointly considering the UAV's trajectory and devices' transmission schedule, the authors in \cite{Z. Wang} use an efficient differential evolution-based method to minimize the maximum energy consumption of all devices in an IoT network. In \cite{B. Liu}, the authors aim to minimize the transmission energy consumption of the sensor nodes within a given data collection time by jointly optimizing the UAV's trajectory and the transmission policy of nodes. In \cite{J. Baek}, the authors consider maximizing the minimum residual energy of sensors after data transmission in order to prolong the network lifetime. The authors in \cite{C. Zhan Y. Zeng} jointly optimize the sensor nodes' wake-up schedule and the UAV trajectory to reduce the maximum energy consumption of all sensor nodes.

     All the above mentioned works only focus on minimizing the energy consumption of ground devices in an UAV-aided wireless network. In contrast, other works consider UAV-related energy consumption minimization when UAVs are deployed in wireless networks. In \cite{Q. Song}, the authors study the problem of minimizing the completion time and the energy consumption of an UAV flying over a large area and propose a fly-and-communicate protocol. The authors in \cite{Y. Zeng} aim to minimize the total energy consumption of the UAV, including both the propulsion and communication energy, in a UAV-enabled system serving multiple ground nodes. The authors in \cite{C. H. Liu} study methods to control a group of UAVs for effectively covering a large geographical region while minimizing their energy consumption. In \cite{D. H. Tran}, the authors minimize the total UAV's energy consumption for a given path by optimizing its velocity.

     Against the above literature, {we consider to minimize the total energy consumption} of the ground network and the UAV by designing an energy efficient UAV trajectory in a cluster-based IoT network. An important difference between our work and existing studies is how the UAV interacts with the ground network. Specifically, existing studies consider the scenario that the UAV directly communicates with each device of the ground network. Such a scenario leads to high energy consumption of both the UAV and ground devices, especially when the network size increases. In contrast, we consider a clustered IoT network and that the UAV only communicates with the cluster head (CH) of each cluster in order to reduce the energy consumption. As such, the UAV trajectory optimization problem is formulated to jointly select CHs and plan the UAV's visiting order to these CHs to minimize the overall energy consumption of the UAV-aided IoT network. {The formulated optimization problem turns out to be one canonical example of combinatorial optimization problems, i.e., the generalized traveling salesman problem (GTSP).}

     In  general,  existing solutions for energy-efficient UAV trajectory planning can be classified into two categories: \emph{traditional methods} (including exact algorithms, heuristic or meta-heuristic algorithms, etc.), and \emph{machine learning based techniques}. Although exact algorithms can provide the optimal solutions (through systematic enumeration, mathematical programming, etc), as the size of the optimization problems increases, their computational costs grow and may prohibit practical implementations \cite{D. Rojas-Viloria}. As for heuristic or meta-heuristic algorithms, there is no guarantee that the obtained solutions are close-to-optimal solutions \cite{K. Zhu}. On the other hand, deep reinforcement learning (DRL) techniques have gained remarkable attention in solving the energy efficient UAV trajectory planning problems. For instance, the authors in \cite{B. Zhang} propose a DRL algorithm to design the UAV cruise route in  a smart city environment, where convolutional neural networks (CNNs) are used for feature extraction and the deep Q-network (DQN) is utilized to make decisions. In \cite{S. F. Abedin}, the authors use the DQN with experience replay memory to solve the formulated energy-efficient trajectory optimization problem, while maintaining data freshness. To provide energy-efficient and fair communication service, the authors in \cite{R. Ding} design a DRL algorithm based on deep neural networks (DNNs) and deep deterministic policy gradient (DDPG) to plan the UAV trajectory in a 3D coverage scenario. With the objective of saving energy, the authors in \cite{Y. Yuan} propose a deep stochastic online scheduling algorithm based on two DNNs and the actor-critic to overcome the traditional DRL's limitations in addressing UAV trajectory optimization problem.

     {In designing a machine learning-based algorithm to solve our formulated combinatorial optimization problem, it is useful to require the algorithm to have the following capabilities: scalability, generalization, and automation on variable-length data structures. This stems from the fact that the number of clusters or nodes in the IoT network may not be the same in different data collection tasks over a given region. The scalability means  that the algorithm is not only able to handle IoT networks with small-scale clusters but also scale to IoT networks with large-scale clusters. The generalization means that the machine learning algorithm should also perform well on unseen problems. The automation of the machine learning algorithm is to automatically execute operations of generalization and scalability on new problem instances, without retraining the model or manually modifying its parameters. In other words, once the model is trained by the designed machine learning algorithm in our work, it can automatically produce good solutions to new IoT networks with different number of clusters and different locations of nodes.}

     {Since combinatorial optimization problems, such as TSP, vehicle routing problem (VRP), etc., are often solved as sequence-to-sequence (Seq2Seq) prediction problems in machine learning \cite{N. Mazyavkina}, we require the designed machine learning algorithm to have an excellent ability to learn policy on sequential data as in our formulated combinatorial optimization problem that also can be seen as a sequential problem. Machine learning algorithms, such as CNN, DQN, and DDPG, are inefficient to handle sequential problems where the current element of the sequence depends on historical information from the previous elements of the sequence. This makes it hard for these algorithms to store information of past elements for very long time \cite{T. T. Nguyen}. Recurrent neural networks (RNNs) with long short-term memory (LSTM) are frequently used to process sequential problems because their hidden units can store historical information for long time steps \cite{N. Mazyavkina}. In addition, RNNs are the state-of-the-art neural networks to tackle variable-length sequences, e.g., variable-size data in our problem, by re-using the neural network blocks and parameters at every step of the sequence \cite{Y. Bengio}. Attention mechanism is another technique to process a variable-length sequence by sharing its parameters \cite{Y. Bengio}. Hence, RNNs and the attention mechanism-based Seq2Seq models that are commonly composed of the encoder component and the decoder component are emerging as attractive techniques to tackle variable-size sequential problems and they show promising results in various domains, see e.g. \cite{H. Hu, J. Lu, J. J. Q. Yu, K. Li, R. Solozabal, B. Zhu}. Given that we formulate the UAV trajectory planning problem in an UAV-aided cluster-based IoT network as a GTSP, Seq2Seq model with RNNs and the attention mechanism are the right ingredients for developing an efficient DRL algorithm to solve this challenging problem.}

	 	
	 	The main contributions of this paper are summarized as follows:
	\begin{enumerate}
		\item We formulate the energy consumption minimization problem in the UAV-IoT system by {jointly selecting CHs from a cluster-based IoT network and planning the UAV's trajectory to the selected CHs.}
				
		\item By viewing the formulated UAV trajectory planning problem in the clustered IoT network as a {combinatorial optimization problem}, we propose a Seq2Seq neural network to model and solve the trajectory planning problem. The inputs to the Seq2Seq neural network are all clusters and the UAV's start/end point; while the output of the Seq2Seq is the UAV's trajectory including the set of selected CHs.
		Reinforcement learning (RL) is used to train the parameters of the Seq2Seq in an unsupervised way to produce a close-to-optimal trajectory that ensures the minimum energy consumption in the UAV-IoT system.
		
		\item Extensive simulations demonstrate that the proposed DRL-based method can find the optimal or close-to-optimal UAV's trajectory and outperforms baseline techniques when evaluating both the energy consumption in UAV-IoT system and the algorithms' computation time. In addition, the trained model by our proposed DRL algorithm shows {good abilities of scalability, generalization, and automation} to deal with IoT networks with different numbers of clusters without the need to retrain the model.
		
	\end{enumerate}
	
	The rest of this paper is organized as follows. Section \ref{SecIII} describes the system model and problem formulation. Section \ref{SecIV} explains how deep reinforcement learning can be used to address the UAV's trajectory planning problem considered in our work. Section \ref{SecVI} presents simulation results. Finally, Section \ref{SecVII} concludes the paper.

	\section{System Model and Problem Formulation}\label{SecIII}
\begin{figure}[t!]
		\centering
		\includegraphics[width=1\linewidth]{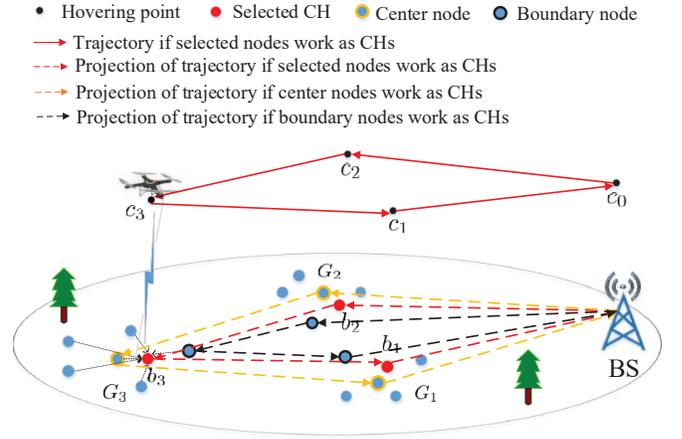}
		\caption{System model of an UAV-aided cluster-based IoT network.}
		\label{trajectory}
	\end{figure}

	We assume that one rotary-wing UAV is dispatched to collect data from $K$ ground clusters. Each cluster $G_k, k =1,\dots, K$, is composed of $N$ nodes, and only one node is selected as the CH, denoted by $b_k, b_k \in G_k$. The selection of CHs will be determined by the proposed algorithm. In each cluster, member nodes are responsible for sensing and collecting the environmental data and then send the collected data to the CH. The UAV is assumed to have the flying-hovering model without considering the acceleration-deceleration pattern. It takes off from the start hovering point $c_0$, corresponding to the ground BS $b_0$, visits each target hovering point $c_k$ in a certain order, which is vertically above each CH $b_k$, and returns to $c_0$ after completing the data collection mission. The location of $b_k$ in the cluster $G_k$ is represented by a 3D Cartesian coordinate $(x_k^{\text{CH}}, y_k^{\text{CH}},0)$, and the position of $n$-th member node in this cluster is $(x_k^{(n)}, y_k^{(n)},0)$. Similarly, the position of each hovering point $c_k$ can be represented as $(x_k^{\text{CH}}, y_k^{\text{CH}}, H)$ where $H$ is the fixed flight height of the UAV. The problem of UAV trajectory planning can be regarded as the determination of hovering points $\{c_k\}_{k=1}^K$ and a permutation of $\{c_k\}_{k=1}^K$ and $c_0$. As an illustrative example in Fig. \ref{trajectory}, if we choose the center node of each cluster as the CH, the energy consumption in the ground network will be minimum because the Euclidean distances between member nodes and their CHs are small in each cluster \cite{A. Ray}. However, this increases the length of the UAV trajectory, and thus increases the energy consumption of the UAV (see the golden dashed line). On the other hand, if a boundary node in each cluster is chosen as the CH, the energy consumption of the UAV will be lower because it has a short trajectory (see the black dashed line). However, in this case the energy consumption for communication in the ground network will increase. Therefore, studying the problem of jointly selecting CHs and planning the UAV's trajectory to minimize the energy consumption of the UAV-IoT network is relevant and very important. This problem will be described in more detail in the following subsections.
	
    \subsection{Channel Model}
     In this work, we consider the air-to-ground channel model as in \cite{J. Yao and N. Ansari} where line-of-sight (LoS) links and nonline-of-sight (NLoS) links are used between the UAV and the ground devices. The probability of a LoS link typically is given by
     \begin{equation}
	    P_{\text{LoS}} = \frac{1}{1+\eta\exp{\left(-\beta[\tau-\eta]\right)}}
	\end{equation}
	where $\eta$ and $\beta$ are environment constants, and $\tau=  \arcsin{(H/d_{k})}180/\pi$, where $d_k = ||c_k-b_k||$ is the distance between the UAV and the ground CH $b_k$ when the UAV hovers at $c_k$.
	The probability of a NLoS link is given by $P_{\text{NLoS}} = 1-P_{\text{LoS}}$.
	The average path loss between $b_k$ and the UAV can be expressed as \cite{J. Yao and N. Ansari}
	\begin{align}
	    \overline{P}_{\text{loss}} = \,& P_{\text{LoS}} \left(10\alpha \log_{10}\left(\frac{4\pi f_c H}{c}\right) + \mu_{\text{LoS}} \right) \nonumber \\
	    & + P_{\text{NLoS}}\left(10\alpha \log_{10}\left(\frac{4\pi f_c H}{c}\right) + \mu_{\text{NLoS}} \right)
	\end{align}
	where $\mu_{\text{LoS}}$ and $\mu_{\text{NLoS}}$ are the average additional losses in LoS and NLoS links, respectively, $\alpha$ is the path loss exponent, $c$ is the speed of light, and $f_c$ is the carrier frequency. Assuming that all CHs have the same transmit power $P_{\text{CH}}$, the average data rate for the communication between each CH and the UAV is defined by \cite{J. Yao and N. Ansari}
	\begin{equation}
	    r_{\text{data}} = B_{\text{width}}\log_2\left(1+\frac{P_{\text{CH}}}{\overline{P}_{\text{loss}}N_{0}}\right)
	\end{equation}
	where $ B_{\text{width}}$ is the communication bandwidth and $N_0$ is the noise power.
	
    \subsection{UAV's Energy Model}
    Without loss of generality, we assume that the UAV flies with a fixed speed $v_\text{UAV}$ from one hovering point to another. The propulsion power consumption of the UAV for movement is given by \cite{M. B. Ghorbel, J. Yao and N. Ansari}
    \begin{align}
	    \label{pf}
	    P_{\text{move}} = \sqrt{\frac{\left(m_{\text{tot}}g \right)^3}{2 \pi r^2_{p}n_{p}\rho}} + \frac{P_{\text{full}}-P_{\text{s}}}{v_{\text{full}}}v_{\text{UAV}} + P_{\text{s}}
	\end{align}
	where $g$, $m_{\text{tot}}$, $r_p$, $n_p$, and $\rho$ are the earth gravity, UAV's mass, propeller radius, number of propellers, and air density, respectively. $P_{\text{full}}$ and $P_\text{s}$ are the hardware power levels when the UAV is moving at full speed $v_{\text{full}}$ and when the UAV hovers, respectively. When the UAV hovers at the hovering point $c_k$ to collect data from the ground CH $b_k$, its power consumption $ P_{\text{hover}}$ for hovering status is obtained by substituting $v_{\text{UAV}}=0$ in (\ref{pf}).
    We assume that the hovering time of the UAV is equal to the data transmission time. Hence, its energy consumption is given by
    \begin{align}
	    E_{c_k} = \frac{D_k}{r_{\text{data}}}(P_{\text{hover}} + P_{\text{com}})
	\end{align}
	where $D_k$ is the amount of data needed to be collected, and $P_{\text{com}}$ is the UAV's communication power. The energy consumption of the UAV for moving from point $c_k$ to another point $c_j$ is given by
	 \begin{eqnarray}
	 \label{eck}
            E_{c_k,c_j} = \frac{||c_k-c_j||}{v_\text{UAV}}P_{\text{move}}.
    \end{eqnarray}
    Hence, {by substituting (\ref{pf}) into (\ref{eck}), the total energy consumption of the UAV in flight can be written as
    \begin{align}\label{eq:E_flight}
        E_{\text{flight}} &= \sum_{k=0}^{K} \sum_{\substack{j=0 \\ j \neq k}}^{K} E_{c_{k},c_{j}}L_{c_{k}} \nonumber \\
         & = \sum_{k=0}^{K} \sum_{\substack{j=0 \\ j \neq k}}^{K}\frac{L_{c_{k},c_{j}}||c_k-c_j|| \left(P_{\text{full}}-P_{\text{s}} \right)}{v_\text{full}} \nonumber \\
         & + \sum_{k=0}^{K} \sum_{\substack{j=0 \\ j \neq k}}^{K}\frac{L_{c_{k},c_{j}}||c_k-c_j||}{v_\text{UAV}} \left(\sqrt{\frac{\left(m_{\text{tot}}g \right)^3}{2 \pi r^2_{p}n_{p}\rho}} + P_{\text{s}} \right), \nonumber \\
         &\quad \forall{c_k,c_j} \in {\mathcal{C}}
    \end{align}} where $\mathcal{C} = \{c_0,c_1,\dots, c_K\}$, $c_k$ is determined by $b_k$, $b_k \in G_k$, and  $L_{c_k,c_j}$ indicates whether the UAV travels from $c_k$ to $c_j$. Specifically, it is defined as
	\begin{equation}
	\label{visit}
	L_{c_k,c_j} = \left\{ \begin{array}{ll}
	1,& \text{the path goes from $c_k$ to $c_j$}\\
	0,& \text{otherwise}.
	\end{array} \right.
	\end{equation}
	As can be seen from \eqref{eq:E_flight}, $E_{\text{flight}}$ is inversely proportional to the speed $v_{\text{UAV}}$. Furthermore, one can show that the choices of $v_{\text{UAV}}$ and hovering points in $\mathcal{C}$ have independent effects on $E_{\text{flight}}$. This means that the UAV speed $v_{\text{UAV}}$ and the hovering points in  $\mathcal{C}$ can be optimized separately. It is simple to see that, according to our system model, in order to minimize $E_{\text{flight}}$, and hence, the overall energy consumption, $v_{\text{UAV}}$ should be set to the maximum flight speed $v_{\text{full}}$. It should be pointed out, however, that for other power models of UAVs (see \cite{Y. Zeng} for example), the optimal value of $v_{\text{UAV}}$ can be any value lower than or equal to $v_{\text{full}}$.

    The following constraints need to be considered for the UAV's trajectory:
	\begin{equation}
	\label{in}
	\sum_{\substack{k=0 \\ k \neq j}}^{K}L_{c_k,c_j} = 1, \quad \forall{c_k, c_j} \in {\mathcal{C}}
	\end{equation}
	\begin{equation}
	\label{out}
	\sum_{\substack{j=0 \\ j \neq k}}^{K}L_{c_k,c_j} = 1, \quad \forall{c_k, c_j} \in {\mathcal{C}}
	\end{equation}
	\begin{equation}
	\label{single}
	\sum_{c_k \in {\mathcal{F}}}\sum_{c_j \in \mathcal{F}}L_{c_k,c_j} \leq |\mathcal{F}| - 1,\quad \forall{\mathcal{F}} \subset \mathcal{C}; |\mathcal{F}| \geq 2.
	\end{equation}
	The constraints (\ref{in}) and (\ref{out}) guarantee that the UAV should visit each point in $\mathcal{C}$ exactly once. Constraint (\ref{single}) enforces that there is only one single trajectory without partial loop exists, where $\mathcal{F}$ is the subset of $\mathcal{C}$ \cite{R. Roberti}.

    \subsection{IoT Network's Energy Models}

	The total energy consumption in the ground network includes energy consumption for intra-cluster communication and energy consumed for data transmission from CHs to the UAV. We use the first-order radio model \cite{W. R. Heinzelman} to calculate the intra-cluster energy consumption. In order to transmit an $l$-bit message to its CH $b_k$, the energy consumed by a member node $n$ is given by \cite{W. R. Heinzelman}
	\begin{equation}
	\label{groundenergy}
	E_{n}^{b_k} = lE_{\text{elec}} + l\left(\chi \varepsilon_\text{fs}d_{n,b_k}^2 + \left(1 - \chi \right) \varepsilon_\text{mp}d_{n,b_k}^4 \right)
	\end{equation}
	where
	\begin{equation}
	\chi = \left\{ \begin{array}{ll}
	1, & d_{n,b_k} \leq d_0 \\
	0, & d_{n,b_k} > d_0
	\end{array} \right.
	\end{equation}
	and
	\begin{equation}
	d_0 = \sqrt{\frac{\varepsilon_\text{fs}}{\varepsilon_\text{mp}}};
	\end{equation}
	$E_{\text{elec}}$ is the dissipated energy per bit in the circuitry, $d_{n,b_k}$ is the distance between CH $b_k$ and member nodes $n$, $d_0$ is the distance threshold, $\varepsilon_\text{fs}$ and $\varepsilon_\text{mp}$ are the radio amplifier's energy parameters corresponding to the free space and multi-path fading models, respectively \cite{W. Heinzelman dissertstion}. On the other hand, the energy consumption of CH $b_k$ to receive an $l$-bit message from member node $n$ is calculated as\cite{W. R. Heinzelman}
	\begin{equation}
	E_{b_k}^{(n)} = lE_{\text{elec}}.
	\end{equation}
	Furthermore, the energy consumed by CH $b_k$ to complete data transmission to the UAV is
	\begin{align}
	   E_{b_k} &= P_{\text{CH}}\frac{D_k}{r_{\text{data}}},
	\end{align}
where $D_k = (N-1)l$.
	
	\subsection{Problem Formulation for UAV's Trajectory}
	Based on the discussed energy models, after the UAV completes a round of data collection task, the total weighted energy consumption of the UAV-IoT system is formulated as
	\begin{align}
	\label{eq19}
	    E\left(b_0,b_1,\dots,b_K \right)
		&=\omega \left(\sum_{k=1}^{K}\sum_{\substack{n=1}}^{N-1}\left(E^{b_k}_n + E_{b_k}^{(n)}\right) + \sum_{k=1}^{K}E_{b_k}\right) \nonumber \\
		&+ (1-\omega) \left(E_{\text{flight}} + \sum_{k=1}^{K}E_{c_k}\right), \,0 \leq \omega \leq 1
	\end{align}
	where $\omega$ is a weighting coefficient that adjusts the energy consumption trade-off between the UAV and the ground networks. Note that the first term is the total energy consumption of the ground network, {which only depends on the positions of CHs,} and the second term is the total energy consumption of the UAV. {$E_{c_k}$ depends on CHs, and $E_{\text{flight}}$ is related to CHs and the visiting order to CHs.}
	With the aim of minimizing the overall weighted energy consumption of the UAV-IoT system, we formulate the optimization problem as jointly selecting CHs and designing the UAV's trajectory, which can be written as
	\begin{eqnarray}
	\label{objectivefunction}
	&&\min_{\substack{\{b_0,b_1,\dots,b_k,\dots, b_K\}\\ b_k\in G_{k}}} \quad  E\left(b_0,b_1,\dots,b_K \right) \nonumber\\
	\\
	&&\text{s.t.} \, (\ref{visit})-(\ref{single}).\nonumber
	\end{eqnarray}

Clearly, the above formulated problem is GTSP,  where the UAV is required to find a tour with the minimal energy consumption of the UAV-IoT system that includes exactly one node from each cluster. Due to the NP-hardness of the formulated problem, it is difficult to solve with conventional methods such as heuristic algorithms. Recent major breakthroughs in DRL have shown that DRL has the ability to successfully solve some combinatorial optimization problems \cite{I. Bello}. Hence, {we propose a sequential model-based DRL method to tackle the problem of jointly selecting CHs and planning the UAV's trajectory.}

	\section{Deep Reinforcement Learning for UAV Trajectory}\label{SecIV}
    \subsection{UAV's Trajectory as Sequence Prediction}

     \begin{figure*}[t]
		\centering
		\includegraphics[width=0.9\linewidth]{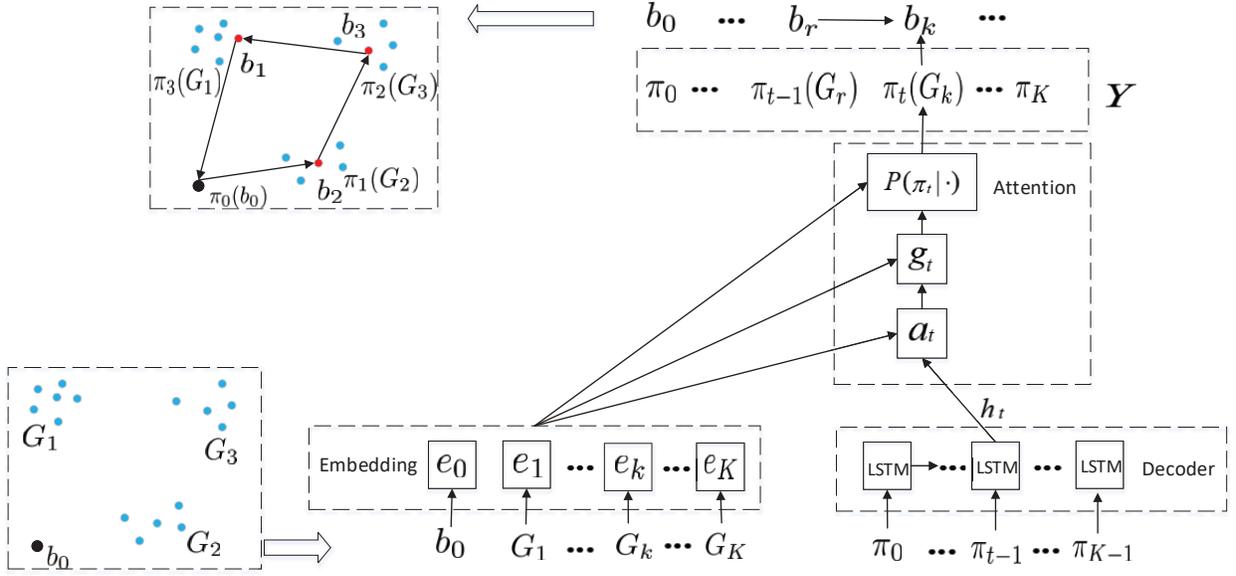}
		\caption{Seq2Seq model with encoder-decoder framework.}
		\label{actorstru}
	\end{figure*}
	
	Because the UAV needs to visit all clusters sequentially to collect data, the trajectory planning problem can be viewed as the visiting decision problem by a policy. This policy can be modeled as a Seq2Seq neural network where one network encodes the input clusters, and then another network is used to convert the encoded information to a visiting order of clusters as its output. Given the start position $b_0$ and $K$ clusters, {the input of the Seq2Seq model is} $\bm{{C}} = \{b_0,G_1, \dots, G_K\}$ and the output is the UAV's visiting order to these elements in $\bm{{C}}$, denoted as $\bm{Y} = \{\pi_0,\pi_1,\dots,\pi_K\}$. Because the UAV takes off from $b_0$, $b_0$ should be in the first position of $\bm{Y}$. For keeping consistency of symbols, we use $\pi_0$ to represent $b_0$ in $\bm{Y}$. Hence, the probability of $\bm{Y}$, i.e., the probability that the UAV follows the corresponding trajectory, can be decomposed using the chain rule as follows
    \begin{equation}
    \begin{aligned}
    \label{chainrule}
        P_\theta(\bm{Y}|\bm{C}) & = \prod_{t=0}^{K}P(\pi_t|\pi_0,\dots,\pi_{t-1}, \bm{C})
    \end{aligned}
    \end{equation}
	where $t$ is the time step, $P(\pi_t|\cdot)$ models the probability of any cluster being visited at the $t$-th time step based on the clusters that have been visited at previous time steps and $\bm{C}$ \cite{I. Bello etc.}. Note that the \emph{stochastic policy} $P_\theta(\bm{Y}|\bm{C})$ is parameterized by $\theta$. In the following subsections, we will use the Seq2Seq neural network architecture in \cite{M. Nazari} to calculate the probability $P(\pi_t|\cdot)$.

    \subsection{Encoder-Decoder Framework for UAV's Trajectory}
     A typical Seq2Seq neural network includes an encoder and a decoder, where the encoder reads and arranges the input sequence into a vector, and the decoder outputs a target sequence by decoding this vector \cite{O. Vinyals}.
    		
    \subsubsection{Encoder}
     Since the inputs are coordinates of the nodes of all clusters, which do not convey sequential information, we use a set of embeddings corresponding to different elements of the input as the encoder in our model instead of using RNNs. Specifically, the embeddings are to map the low-dimensional inputs to a high $D$-dimensional vector space. By doing so, the computational complexity of the embedding layer is reduced without reducing its efficiency. The mapping from the input $\bm{C}$ to the embedded output $\overline{\bm{C}}$ is shown as
    \begin{equation}
        \overline{\bm{C}} = W_b\bm{C}
    \end{equation}
    where $W_b$ is the embedding matrix, $\overline{\bm{C}} = \{e_k\}_{k=0}^K$, $e_k\in\mathbb{R}^D$. For example, in Fig. \ref{actorstru}, there are three clusters and one start position, hence $\bm{C} = \{b_0, G_1, G_2, G_3\}$. After embedding, $\bm{C}$ is converted into  $\overline{\bm{C}} = \{e_0, e_1, e_2, e_3\}$.

    \subsubsection{Decoder}
    Since the hidden units of the RNN can be used for learning historical information, it is very common in the literature to use the RNN as the decoder in the encoder-decoder framework. However, the traditional RNN shows poor performance in dealing with the problem of the long-term dependencies, which makes it difficult to be trained in practice \cite{Y. Zhu}. Hence, we use the LSTM which is capable of learning long-term dependencies to construct a RNN as the decoder. The number of decoding steps is equal to the length of $\overline{\bm{C}}$. At each decoding step $t$, the hidden state $h_t\in \mathbb{R}^{D}$ of the LSTM, which stores information of previous steps, and the embedded $\overline{C}$ are used to generate the conditional probability $P(\pi_t|\pi_0,\dots,\pi_{t-1}, \bm{C})$ for deciding the output in this step. Calculating the conditional probabilities is performed by the attention mechanism, whose details are described next.

    \subsubsection{Attention Mechanism}
    Attention mechanism is used to improve the encoder-decoder model, which allows the model to give different weights to different elements of the input \cite{O. Vinyals A. Toshev}. For planning of the UAV's trajectory, attention mechanism tells us the relationship between each cluster in $\bm{C}$ at current step $t$ and the output $\pi_{t-1}$ of the last decoding step. The most relevant cluster with the maximum probability is chosen at decoding step $t$. Specifically, $h_t$ is the hidden state of the LSTM at decoding step $t$. The quantity $a_t^{k}$ represents how relevant each element $e_k$ in $\overline{\bm{C}}$ is at decoding step $t$. It is calculated using the softmax function\footnote{The softmax function is defined as: $a_t^{k} = \frac{\exp\{u_t^{k}\}}{\sum_{j=0}^K\exp\{u_t^{j}\}}$.} as
    \begin{equation}
    \label{akt}
        a_t^k = \text{softmax}\left(u_t^k\right)
    \end{equation}
    where
    \begin{equation}
    \label{u_k}
        u_t^k = \varphi_a \tanh\left(W_1 e_k + W_2 h_t\right),
    \end{equation}
    with $\varphi_a\in \mathbb{R}^{1\times D}$, $W_1\in \mathbb{R}^{D \times D}$, $W_2\in \mathbb{R}^{D \times D}$.
    The context vector $g_t\in \mathbb{R}^D$ is computed as
    \begin{equation}
        g_t = \sum_{k=0}^{K} a_t^k e_{k}.
    \end{equation}
	Then, we combine $g_t$ with the embedded inputs
	\begin{align}
	   \label{ukt}
	   \widetilde{u}_t^k = \varphi_g\tanh{}\left(W_3 e_k + W_4 g_t\right)
	\end{align}
	with $\varphi_g\in \mathbb{R}^{1 \times D}$, $W_3\in \mathbb{R}^{D \times D}$, $W_4\in \mathbb{R}^{D \times D}$. The vector $\widetilde{u}_t = \{\widetilde{u}_t^0, \widetilde{u}_t^1, \dots, \widetilde{u}_t^k, \dots, \widetilde{u}_t^K \} \in \mathbb{R}^{(K+1)}$ is called the logits. To encourage exploration, we use a logit clipping function to control the distribution of the logits
	\begin{align}
	    \bar{u}_t = C_{\text{L}}\tanh{(\widetilde{u}_t)}
	\end{align}
	where $C_{\text{L}}$ is a hyper-parameter that limits the range of the logits to $[-C_{\text{L}}, C_{\text{L}}]$, and hence, the entropy associated with $P(\cdot)$. The  value of $C_{\text{L}}$ is set to 10 by following \cite{I. Bello}.
	To avoid clusters being visited more than once, we apply the mask vector to $\bar{u}_t$ to mark clusters that have been visited before:
	\begin{align}
	    \widehat{u}_t = \{\widehat{u}_t^0,\dots,\widehat{u}_t^k,\dots, \widehat{u}_t^K \} = \bar{u}_t + \mathcal{M}_t
	\end{align}
	where $\mathcal{M}_t \in \mathbb{R}^{(K+1)}$ is the mask vector, which is initialized to a vector of all zeros and its values are updated at each decoding step. If a cluster is selected for access, we update the value in the corresponding position of $ \mathcal{M}_t$ to $-\infty$.
	Then, the element in the corresponding position of $\widehat{u}_t$ also becomes $-\infty$.	
	Finally, we compute the probability of each element in $\widehat{u}_t$ as
	\begin{align}
	    	P(\pi_{t}|\pi_0,\pi_1,\dots,\pi_{t-1}, \bm{C}) = \text{softmax}\left(\widehat{u}_t\right) \label{u_kp}
	\end{align}
	where the negative infinities in $\widehat{u}_t$ get zeroed out after using the softmax function.  We choose the cluster pointed by the highest probability as the output at decoding step $t$ and update the value of the same position in $ \mathcal{M}_t$ to $-\infty$. Thus, each $P(\cdot)$ distribution is represented by the softmax function over all elements in the input sequence. The learnable variables are $\varphi_a, \varphi_g, W_1, W_2, W_3$, and $W_4$, which make up the policy parameter $\theta$.

    We further give an example to explain how the masking mechanism works. As shown in Fig.~\ref{actorstru}, there are three clusters and one start position; hence, $\bm{C} = \{b_0, G_1, G_2, G_3\}$. After embedding, $\bm{C}$ is converted into  $\overline{\bm{C}} = \{e_0, e_1, e_2, e_3\}$. We assume that the outputs of the decoder network at decoding step 0 and 1 are $b_0$ and $G_2$, respectively. In order to get the output of decoding step 2, $\widetilde{u}_2 = \{\widetilde{u}_2^0, \widetilde{u}_2^1, \widetilde{u}_2^2, \widetilde{u}_2^3 \}$ is obtained by equations (\ref{akt})--(\ref{ukt}), and the mask vector is $ \mathcal{M}_2 = \{-\infty,0,-\infty,0\}$.  By summing the elements of $\bar{u}_2$ and $ \mathcal{M}_2$ at the same position, we can obtain  $\widehat{u}_2 = \{ \widehat{u}_2^0, \widehat{u}_2^1, \widehat{u}_2^2, \widehat{u}_2^3 \} = \{-\infty, \bar{u}_2^1, -\infty, \bar{u}_2^3\}$. Applying equation (\ref{u_kp}) to $\widehat{u}_2$, we assume the final probability distribution over $\widehat{u}_2$ is calculated as $\{0, 0.2, 0,0.8\}$. Hence, the cluster $G_3$ is selected at this step because the highest probability value points to it. Then, the mask vector is updated as $ \mathcal{M}_2 = \{-\infty,0,-\infty,-\infty\}$. As we can see, the masked clusters cannot be visited again. Hence, the masking scheme in our proposed algorithm can effectively prevent the clusters from being visited multiple times.
	
	\subsubsection{Selection of CHs}
	We assume that the output $\pi_t$ of the model at step $t$ is the cluster $G_k$, and its CH $b_k$ is chosen by
	\begin{align}
	\label{chselection}
	    b_k = \text{min}\{E_{(b_r,n)}\}_{n=1}^{N}
	\end{align}
	where $b_r$ is the CH of cluster $G_r$ that is visited at step $(t-1)$, $E_{(b_r,n)}$ is the energy consumption of the UAV and the ground IoT network when the UAV flies from the CH $b_r$ to a node $n$ in the next cluster that will be visited. The node $n$ in $G_k$ that can guarantee the minimum energy consumption of the UAV-IoT from $b_r$ to $n$ is selected as the CH of the cluster $G_k$. In the example of Fig. \ref{actorstru}, the start point $b_0$ is visited at the $0$-th decoding step. Then, the output $\pi_1$ of the decoding step 1 is the cluster $G_2$ because it has the highest probability $P(\pi_1|\pi_0, \bm{C})$. We calculate the overall energy consumption of the UAV-IoT from $b_0$ to each node in $G_2$ and choose the CH by (\ref{chselection}).  Finally, we obtain a set of sorted CHs and output the UAV's trajectory, which is shown as
	\begin{equation*}
	    b_0(\pi_0)\xrightarrow{} b_2(\pi_1)\xrightarrow{} b_3(\pi_2)\xrightarrow{} b_1(\pi_3).
	\end{equation*}
	{The above trajectory may not be the best. Hence, we need to train the policy parameter $\theta$ from samples by RL to produce the optimal or close-to-optimal trajectory.}
	
	\subsection{Training Method}
	 In RL, an agent optimizes its behavior by interacting with the environment. The goal of the agent is to search for an optimal policy that can solve the constrained optimization problem through iterative training. All ground clusters, our objective function, and all constraints are considered as the environment. Note that for the agent, the environment is actually treated as a black box. {The goal of the agent is to maximize the accumulated rewards by learning an optimal policy which is a mapping of states and actions. }
	 {\subsubsection{State} The state of the problem at time step $t$ is composed of the coordinates of all clusters, the location of the UAV, and the energy consumption.}
	 {\subsubsection{Action} The action for the UAV at current step $t$ is the selection of the next cluster and its CH to be visited. Hence, we define the output of the attention mechanism and the CH selection as the action at each step.}
	 {\subsubsection{Reward} The reward function is defined as the negative of the total energy consumption in the formulated problem (\ref{objectivefunction}). The reward of one full episode generated under the policy is denote as $R = -E$.}
	
	 {REINFORCE\cite{R. J. Williams}, the well-known policy gradient, is employed in this paper, and the UAV works as the agent. Unlike value-based methods such as DQN that finds the optimal policy through Q-values, a policy gradient method directly optimizes the policy by changing its parameters.
	 The REINFORCE algorithm uses an estimate of the gradient of the expected reward to update the policy parameter $\theta$. The agent observes a full sequence that includes all states, actions, and rewards from start to finish generated under the policy. We compute the sum reward from this sequence by setting the discount factor to one, which is actually based on the real observed return. To train the proposed Seq2Seq model, the REINFORCE algorithm includes the actor network (policy network) and the critic network (value network).} The Seq2Seq model works as the actor network that generates a set of ordered CHs for a given input problem instance. In the critic network, the output probabilities of the actor network are used to compute a weighted sum of the embedding inputs. Then, the weighted sum vector is fed into two-fully connected layers with one ReLU activation and one linear layer with a single output. The critic network, denoted by $\psi$, provides an approximated baseline of the solution for any problem instance to reduce the variance of gradients\cite{V. R. Konda}. Given a problem instance $\bm{C}$, the training objective is the expected reward, which is defined as
	\begin{equation}
	    \label{J}
	    J\left(\theta|\bm{C}\right) = \mathbb{E}_{\bm{Y}\sim p_{\theta(.|\bm{C})}} [R].
	\end{equation}
    One can use policy gradient and stochastic gradient descent to optimize $\theta$. The gradient of (\ref{J}) is formulated using REINFORCE algorithm,  which can provide an unbiased gradient, as follows
	\begin{equation}
	   \label{gra}
	    \nabla_{\theta}J\left(\theta|\bm{C}\right) = \mathbb{E}_{\bm{Y}\sim p_{\theta(.|\bm{C})}}\left[\left(R - V_{\psi}\left(\bm{C}\right)\right) \nabla_{\theta}\log p_{\theta}\left( \bm{Y} | \bm{C}\right) \right]
	\end{equation}
	where $V_{\psi}\left(\bm{C}\right)$ is a parametric baseline implemented by the critic network { to reduce the variance of the gradient}. { We use batches to speed up the training process. Assuming there are $B$ problem instances in each batch,} the gradient in (\ref{gra}) is approximated {with Monte Carlo sampling as}
	\begin{equation}
	    \nabla_{\theta}J\left(\theta\right) \approx \frac{1}{B} \sum_{i=1}^B \left( R_i-V_{\psi}\left(\bm{C}_i\right) \right) \nabla_{\theta} \log p_{\theta} \left(\bm{Y}_i|\bm{C}_i\right).
	\end{equation}
    The critic network is trained by using stochastic gradient descent on a mean squared error objective between $V_\psi \left(\bm{C}_i \right)$ and the actual energy consumption, which is given by
	\begin{equation}
	    \mathcal{L}(\psi)=\frac{1}{B}\sum_{i=1}^{B}\left(V_{\psi}\left(\bm{C}_i\right)-R_i \right)^2.
	\end{equation}
	
	The training procedure of the actor network and the critic network is shown in Algorithm \ref{alg1}. The parameters of the actor and critic networks are updated iteratively by using the Adam algorithm\cite{D. P. Kingma}.

	\begin{algorithm}[t!]
	\caption{{REINFORCE with the baseline algorithm}}
	\label{alg1}
	\begin{algorithmic}[1]
		\renewcommand{\algorithmicrequire}{\textbf{Input:}}
		\REQUIRE  Batch size $B$, training step $S$, training data set $\bm{\mathcal{Q}} = \{\bm{C}_1,\dots, \bm{C}_{S\times B}\}$
	    \STATE Initialize the actor network parameter $\theta$ and the critic network parameter $\psi$\\
	    \FOR{$s$ = 0 to $S-1$}
	         \STATE Obtain train data $\bm{\mathcal{C}}_s = \{\bm{C}_{1+(s\times B)}, \dots, \bm{C}_{(s\times B) +B}\}$  from $\bm{\mathcal{Q}}$ for the current training step\\
	         \STATE Find CHs and calculate $R_i$ for each $\bm{C}_i$ in $\bm{\mathcal{C}}_s$ with \\the actor network\\
	         \STATE Calculate $V_{\psi}\left(\bm{C}_i \right)$ with the critic network\\
		    \STATE $d\theta \gets \frac{1}{B} \displaystyle \sum_{i=1+(s\times B)}^{(s\times B)+B} \left( R_i-V_{\psi}\left(\bm{C}_i\right) \right) \nabla_{\theta} \log p_{\theta} \left(\bm{Y}_i|\bm{C}_i\right)$ \\
		    \STATE $\mathcal{L}(\psi) \gets \frac{1}{B}\displaystyle \sum_{i=1+(s\times B)}^{(s\times B)+B}\left(V_{\psi}\left(\bm{C}_i \right) - R_i\right)^2$ \\
		    \STATE $\theta \gets \text{Adam}\left(\theta, d\theta\right)$ \\
		    \STATE $\psi \gets \text{Adam}\left(\psi, \nabla_{\psi}\mathcal{L}\psi\right)$ \\
		\ENDFOR
		\RETURN $\theta$\\
	\end{algorithmic}
    \end{algorithm}

\section{Numerical Results}\label{SecVI}

We compare the proposed approach with the following three common baseline methods:
\subsubsection{Greedy}
The greedy algorithm follows the problem-solving heuristic of making the locally optimal choice at each stage\cite{T.H. Cormen}. When looking for a solution, it always takes the best immediate or local decision, which may lead to poor solutions for some problems. The greedy algorithm is usually faster than exact methods because it does not consider the details of possible alternatives.

\subsubsection{Gurobi}

The Gurobi optimizer is the most powerful mathematical optimization solver for linear programming, quadratic programming, mixed integer linear programming, mixed-integer quadratic programming, mixed-integer quadratically constrained programming, etc\cite{gurobi}. It is an exact algorithm solver that enables users to build mathematical models for their problems and produces the optimal solutions globally. For the optimization problem considered in this paper, the presented optimal solutions are obtained using Gurobi.

\subsubsection{{Ant colony optimization (ACO)}}
{ACO is a meta-heuristic method inspired by the observation of real ant colonies, which can be used to solve various combinatorial optimization problems. In ACO, multi-ants leave their nest and walk randomly until they find food. Each ant deposits a substance called pheromone along its trail so that the other ants can follow. An ant tends to choose the path with the highest pheromone concentration because its length is the shortest. Since our formulated problem is GTSP, we use the extended ACO method proposed in \cite{J. Yang} to compare with our proposed DRL technique. In the following simulations, the parameters of ACO are set as follows. The number of ants is 30, the number of iterations is 200, the pheromone evaporation coefficient is set as 0.1, and the importance of pheromone and the relative importance of visibility are 1 and 5, respectively. }

\begin{table}[t!]
	\centering
	\caption{Simulation parameters}
	\label{parameters}
	\scriptsize
	 \begin{tabular}{p{1.3cm}<{\centering}|p{2.2cm}<{\centering}||p{1.3cm}<{\centering}|p{2.3cm}<{\centering}}
		\hline{}
		\textbf{Parameter} & \textbf{Value} & \textbf{Parameter} & \textbf{Value}\\
		\hline\hline
		$P_{\text{CH}}$ & 21 $\text{dBm/Hz}$ {\cite{M. B. Ghorbel}} & $r_p$ & 20 cm {\cite{H. Ghazzai}}\\
		\hline
        $B_{\text{width}}$ & 1 MHz {} & $n_p$ & 4 {\cite{H. Ghazzai}}\\
        \hline
        $N_0$ & $-174$ dBm/Hz {\cite{M. B. Ghorbel}} & $\rho$ & 1.225 kg/$\text{m}^3$ {\cite{H. Ghazzai}}\\
        \hline
        $f_c$ & 2 GHz {\cite{M. B. Ghorbel}} & $m_{\text{tot}}$ & 500 g {\cite{H. Ghazzai}}\\
        \hline
        $\alpha$ & 3  {\cite{M. B. Ghorbel}} & $P_{\text{full}}$  & 5 W {\cite{H. Ghazzai}}\\
        \hline
        $H$ & 50 m  & $P_{\text{s}}$ & 0 W {\cite{H. Ghazzai}}\\
        \hline
        $\mu_{\text{LoS}}$, $\mu_{\text{NLoS}}$ & 1 $\text{dB}$, 20 $\text{dB}$ {\cite{A. Al-Hourani}} &$v_{\text{UAV}}=v_{\text{full}}$ & 15 $\text{m/s}$ {\cite{M. B. Ghorbel}} \\
        \hline
        $\beta$ &  0.03  {\cite{M. B. Ghorbel}} & $P_{\text{com}}$ &  0.0126 W {\cite{M. B. Ghorbel}}\\
        \hline
        $\eta$ & 10  {\cite{M. B. Ghorbel}} & $N$ & 20\\
        \hline
        $\varepsilon_\text{fs}$ & {10 $\text{pJ/bit/m}^2$ \cite{W. R. Heinzelman}} & $\varepsilon_\text{mp}$ & {0.0013 $\text{pJ/bit/m}^2$ \cite{W. R. Heinzelman}}\\
        \hline
        {$l$} & {1 MB} &  & \\
        \hline
		\end{tabular}
\end{table}

To thoroughly evaluate the performance of the proposed DRL algorithm, we compare the trajectories of the UAV and the energy consumptions obtained by the proposed algorithm with that obtained by three baseline methods. It should be noted that the energy consumption mentioned in all comparisons refers to the energy consumption in one communication round. In each round, member nodes send data to their CHs, and the UAV visits these CHs to collect data.

{\subsection{Complexity Comparison}
In terms of computational complexity, the greedy algorithm has $O(KN)$ time complexity, which performs $K$ steps to visit all clusters and consumes $O(N)$ operations to select a CH at each step. The computational complexity of ACO is estimated by $O(I_{\text{max}}K^2M_{\text{ant}}N)$ where $I_{\text{max}}$ is the number of iterations and $M_{\text{ant}}$ is the number of ants \cite{J. Yang}. At inference, the computational complexity of the attention mechanism in our proposed algorithm is $O(K+1)$ at each decoding step, and the computational complexity of selecting a CH is $O(N)$. We have to perform $K+1$ steps to output the final result, and hence, the total computational complexity of our proposed algorithm can be further simplified as $O((K+1)^2+KN)) \approx O(K(K+N))$, which is lower than the complexity of the ACO.  According to Gurobi's website \cite{gurobi}, they use the branch-and-bound method to solve optimization problems. Thus, the computational complexity of Gurobi is ultimately exponential, which is worse than our proposed algorithm.}

\subsection{Environment and Parameters Settings}

\begin{figure}[!t]
	\centering
	\includegraphics[width=1\linewidth]{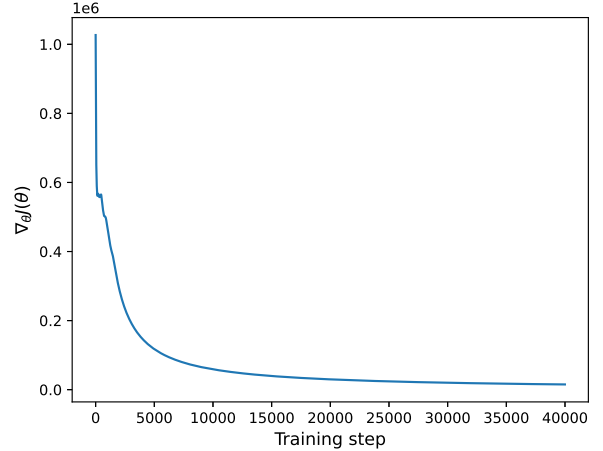}
	\caption{{Training curve of the actor network.}}
	\label{loss}
\end{figure}

We consider a target area of $1\, \text{km} \times 1 \,\text{km}$ where the BS is located with coordinate $b_0 = (500\, \text{m}, 0 \,\text{m})$. Simulation parameters are presented in Table \ref{parameters}. We employ Pytorch 1.4 and Python 3.7 on a computer with 1 NVIDIA TESLA P100 GPU to implement the proposed DRL algorithm. {Each problem instance $\bm{C}$ is composed of the initial location $b_0$ and $K$ clusters. The center $(x_k, y_k)$ of each cluster $G_k$ is firstly sampled from the distribution $ torch.rand(2, 1) * 1000$. Then, the nodes in each cluster are sampled from the uniform distribution $G_k = np.random.uniform([x_k-\zeta, y_k-\zeta], [x_k+\zeta, y_k+\zeta], [N, 2])$, where $\zeta$ is a constant, $[x_k-\zeta, y_k-\zeta]$ represents the left and lower boundaries of cluster $G_k$ in the 2-dimensional space, $[x_k+\zeta, y_k+\zeta]$ is the right and upper boundaries, $0 < x_k-\zeta < x_k+\zeta < 1000$, $0 < y_k-\zeta < y_k+\zeta < 1000$, and all clusters do not overlap with each other, i.e., $G_1 \cap \dots \cap G_k \cap \dots \cap G_K = \emptyset$. All sampled problem instances form the final training data set $\bm{\mathcal{Q}}$.} Setting $K=4$, we implement 40,000 training steps to train the model where the batch size $B$ is equal to 256 at each training step. Each element in any problem instance $\bm{C}$ is embedded into a vector of size 128 by the encoder network. Accordingly, we use LSTM cells with 128 hidden units in the decoder network. The initial learning rate of the actor network and the critic network is set at 0.0001.

{Fig. \ref{loss} shows the training curve of the actor network. One can see that the value of $\nabla_{\theta}J\left(\theta\right)$ decreases sharply in early steps, which is due to the rough approximation at initialization that causes a large loss. When the number of iterations increases,
$\nabla_{\theta}J\left(\theta\right)$ stabilizes and the proposed algorithm converges.}

\begin{figure*}[!t]
  \centering
    \subfigure[]{\includegraphics[width=0.5\textwidth]{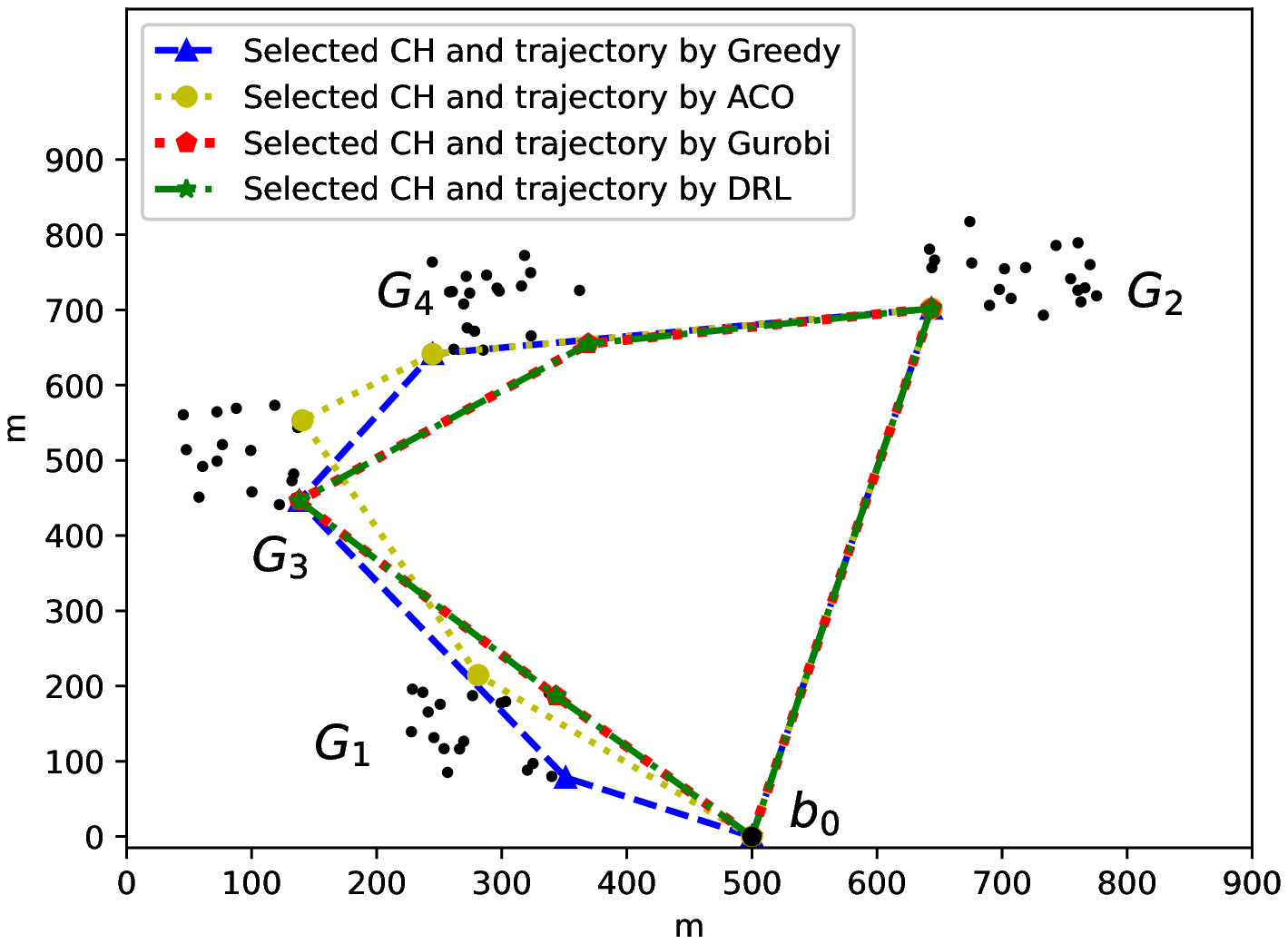}}\hspace{-0.2in}
	\subfigure[]{\includegraphics[width=0.5\textwidth]{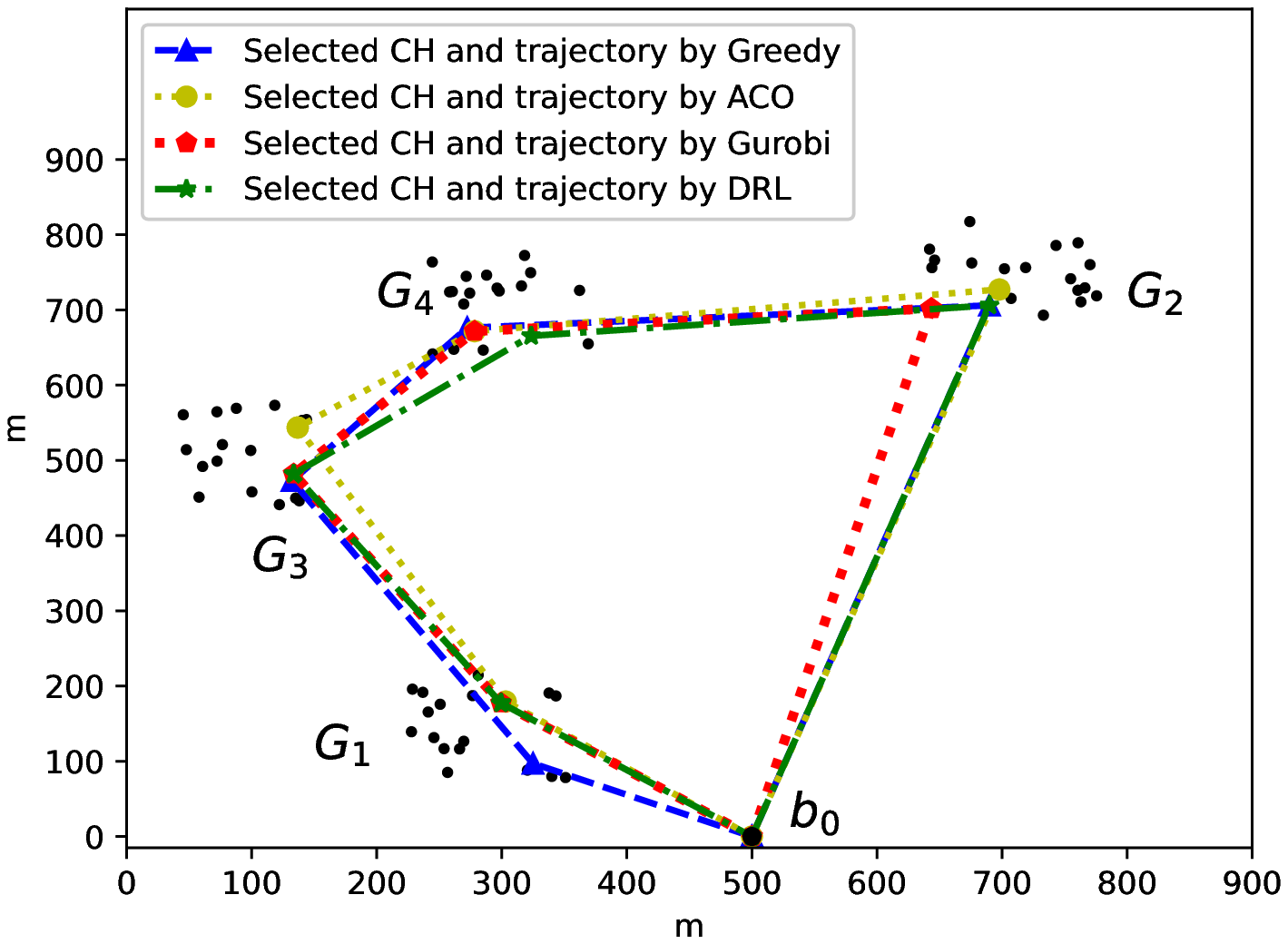}} \\

    \subfigure[]{\includegraphics[width=0.5\textwidth]{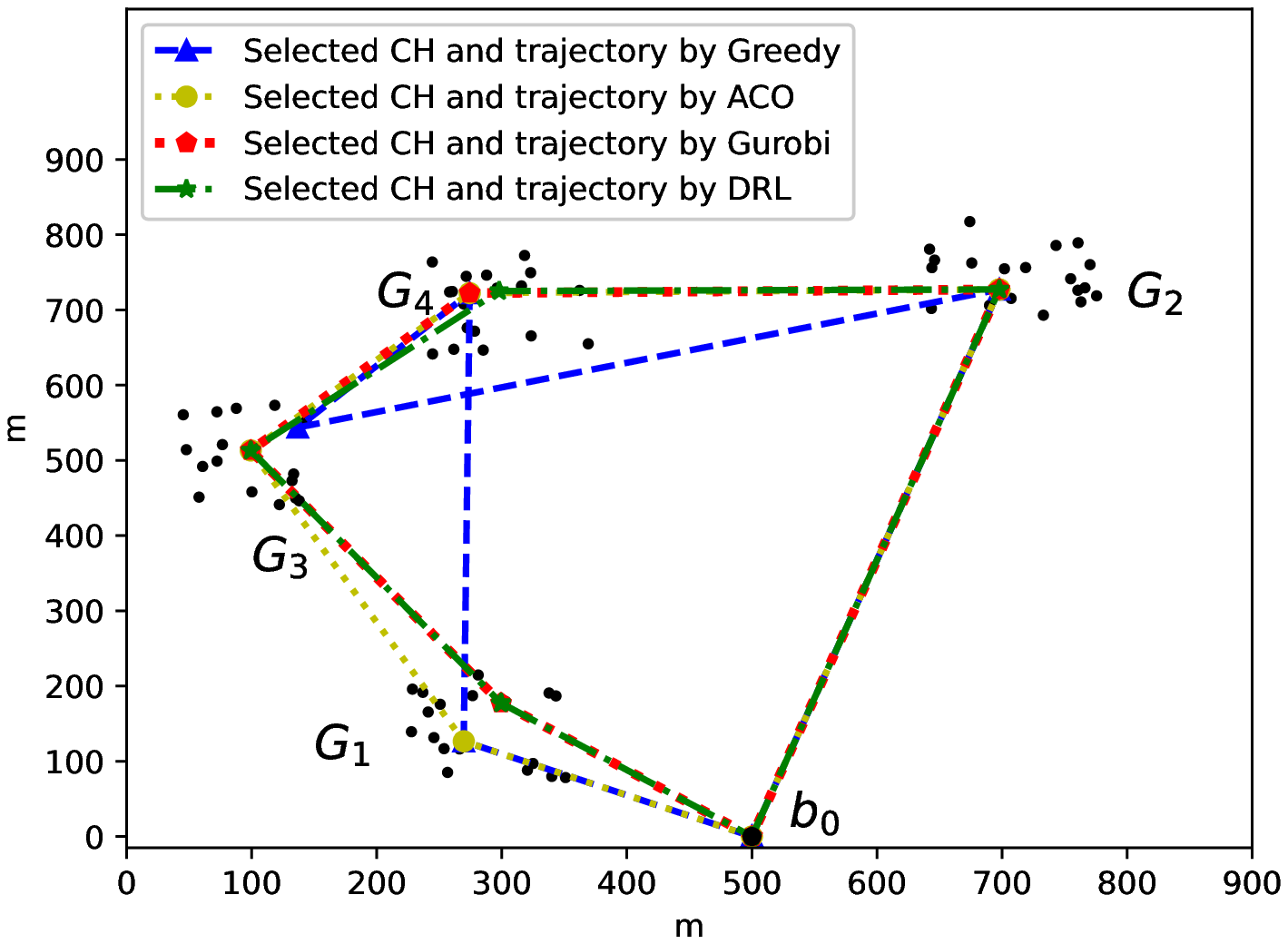}}\hspace{-0.2in}
	\subfigure[]{\includegraphics[width=0.5\textwidth]{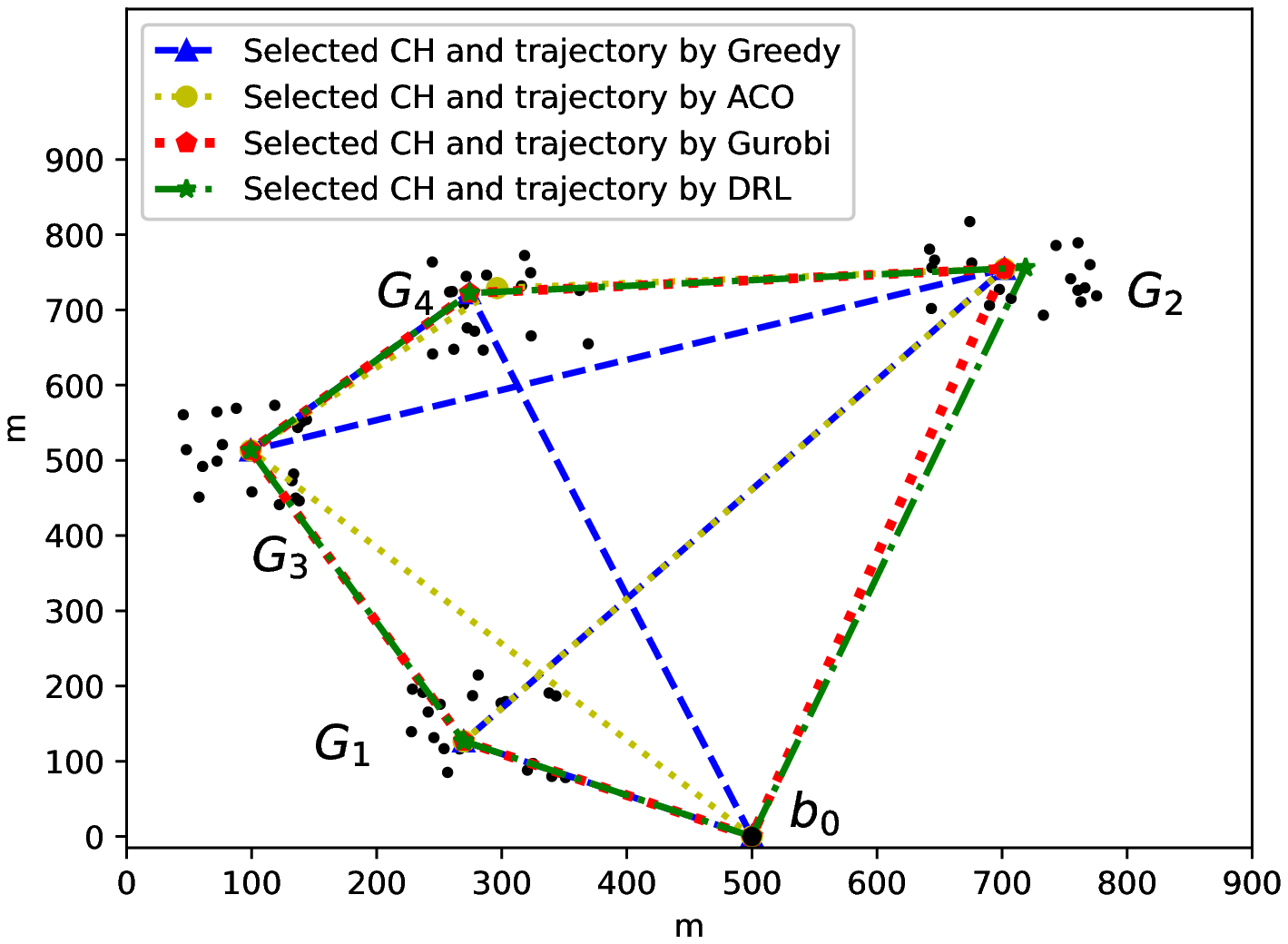}} \\
     \caption{{Trajectory comparison of DRL, greedy algorithm, ACO,  and Gurobi for different values of $\omega$. (a) $\omega = 0$. (b) $\omega = 0.3$. (c) $\omega = 0.6$. (d) $\omega = 0.9$.}}
	 \label{trajectory comparison}
	\vspace{0in}
\end{figure*}

\begin{figure}[!t]
	\centering
	\includegraphics[width=1\linewidth]{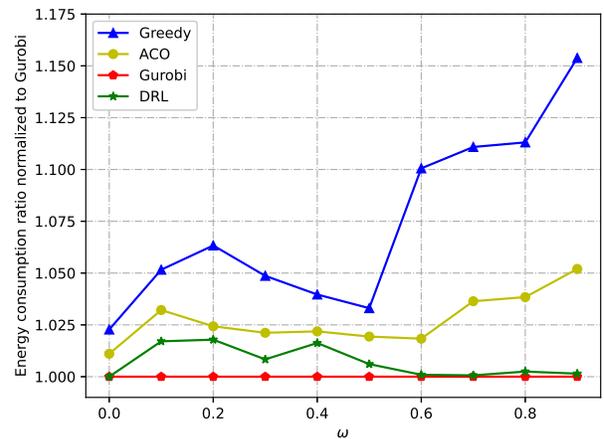}
	\caption{{Energy consumption comparison for 4 clusters.}}
	\label{comparisonratio}
\end{figure}

\subsection{Trajectory and Energy Consumption Comparison}

To show the effectiveness of the proposed algorithm, we compare its performance with the performances of the greedy algorithm, ACO, and Gurobi in this section. We generate a problem instance with four clusters $\{G_1, G_2, G_3, G_4\}$ in the same way as generating the train data. Then, the test data is fed into the trained model to evaluate the proposed DRL algorithm.

In Fig. \ref{trajectory comparison} (a), the value of $\omega$ in (\ref{eq19}) is set to $0$, which means that the goal is to minimize the energy consumption of UAV only, which is proportional to UAV's flight distance in our system model. Since the greedy algorithm only yields locally optimal solutions by visiting the nearest next CH as shown in  Fig. \ref{trajectory comparison} (a), it will not achieve the shortest UAV trajectory. However, the trajectory generated by our proposed DRL algorithm completely coincides with the optimal one obtained from Gurobi, which ensures the energy consumption of the UAV is minimum. In addition, there is a visible gap between the trajectory generated by ACO and the optimal trajectory.

For the results in Fig. \ref{trajectory comparison} (b), we set $\omega=0.3$, which means that the energy consumptions of ground nodes and UAV account for $30\%$ and $70\%$ of the total energy consumption, respectively. In order to minimize the total energy consumption in this case, the CH of each cluster should be between the center and the edge of the cluster and close to the edge. Obviously, all four algorithms can select CHs in the right position as well as plan trajectories to access these CHs. However, the trajectory obtained by our proposed DRL algorithm and the one by Gurobi are almost identical, which exhibits our proposed algorithm can produce the close-to-optimal solution.

The results in Fig. \ref{trajectory comparison} (c) are obtained by setting $\omega = 0.6$, which means that the energy consumption of the ground nodes accounts for a larger proportion of the total energy consumption. As a consequence, it is expected that CHs should be closer to the center of the cluster. The visiting path produced by the greedy algorithm is
\begin{equation*}
    b_0 \rightarrow G_{1} \rightarrow G_{4}  \rightarrow G_{3} \rightarrow G_{2} \rightarrow b_0
\end{equation*}
which travels suitable CHs, but does not present the optimal path to access these CHs. The trajectory produced by ACO is much better than the one of the greedy algorithm, which is given by
\begin{equation*}
    b_0 \rightarrow G_{1} \rightarrow G_{3}  \rightarrow G_{4} \rightarrow G_{2} \rightarrow b_0.
\end{equation*}
However, our algorithm not only can find the appropriate CHs but also plan the optimal access path to these CHs. The trajectories found by the proposed DRL algorithm and Gurobi almost coincide again, as can be seen in Fig. \ref{trajectory comparison} (c).

\begin{figure*}[!t]
	\centering
	 \subfigure[]{\includegraphics[width=0.5\textwidth]{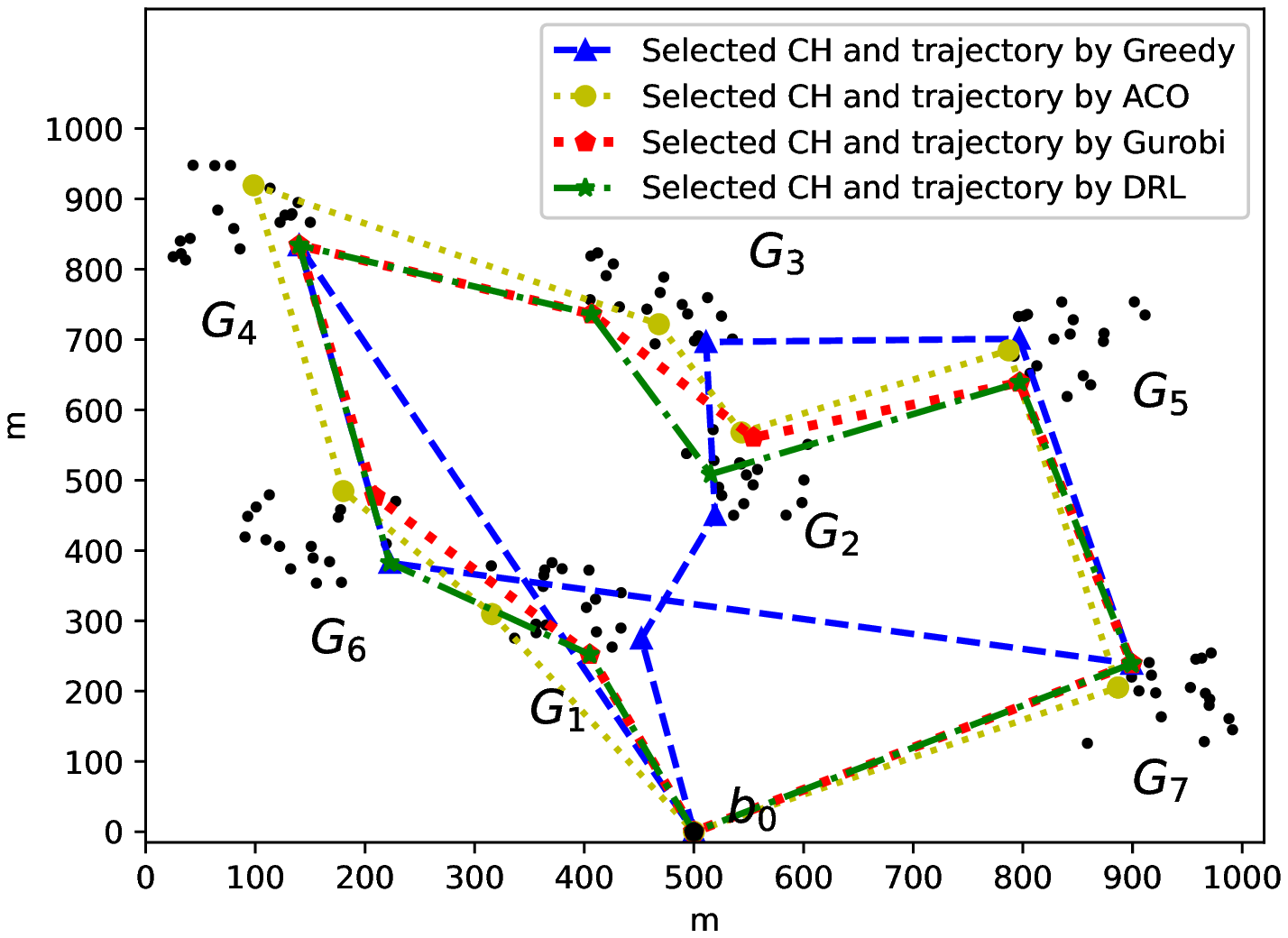}}\hspace{-0.2in}
	\subfigure[]{\includegraphics[width=0.5\textwidth]{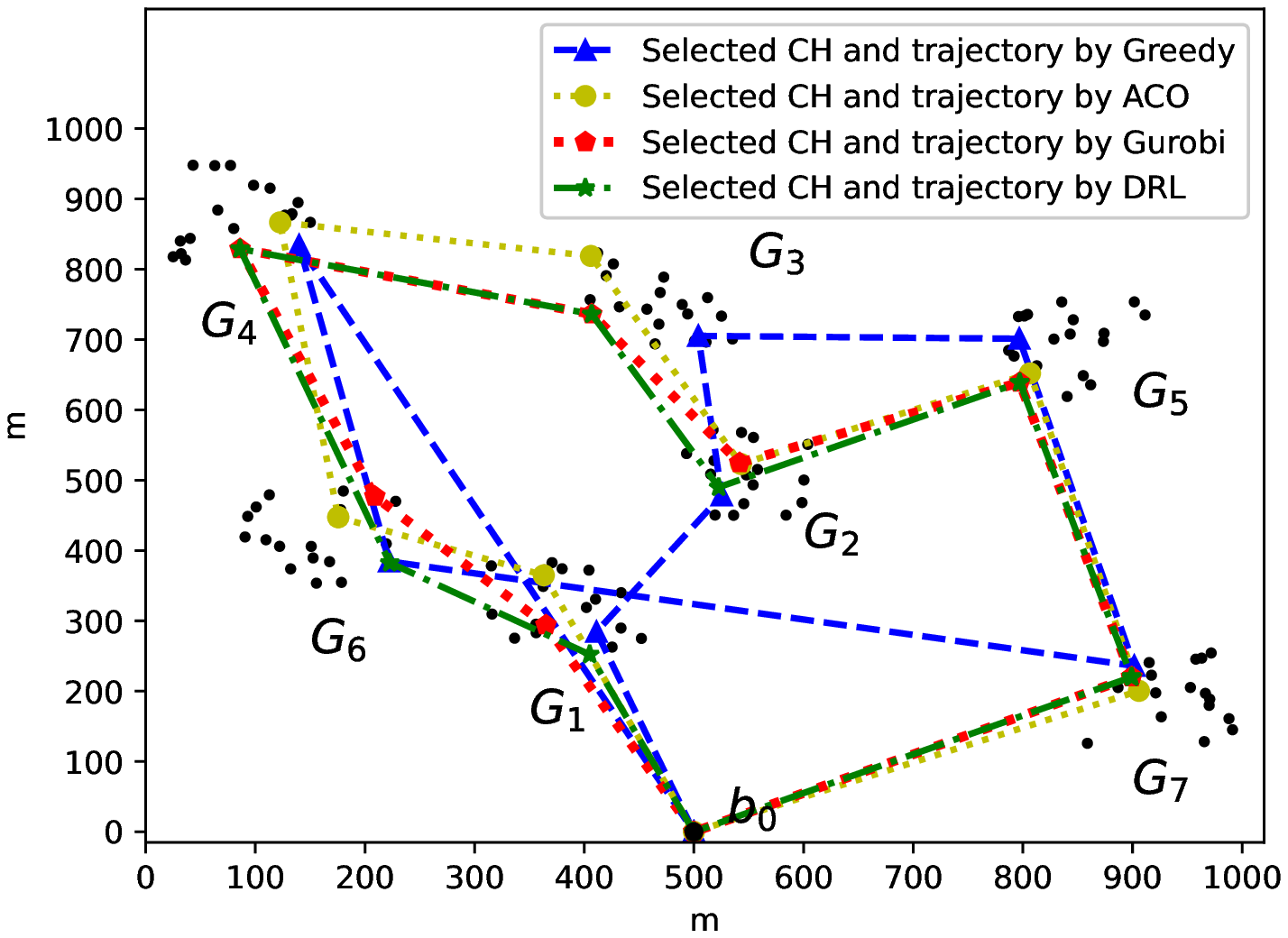}} \\
	
	 \subfigure[]{\includegraphics[width=0.5\textwidth]{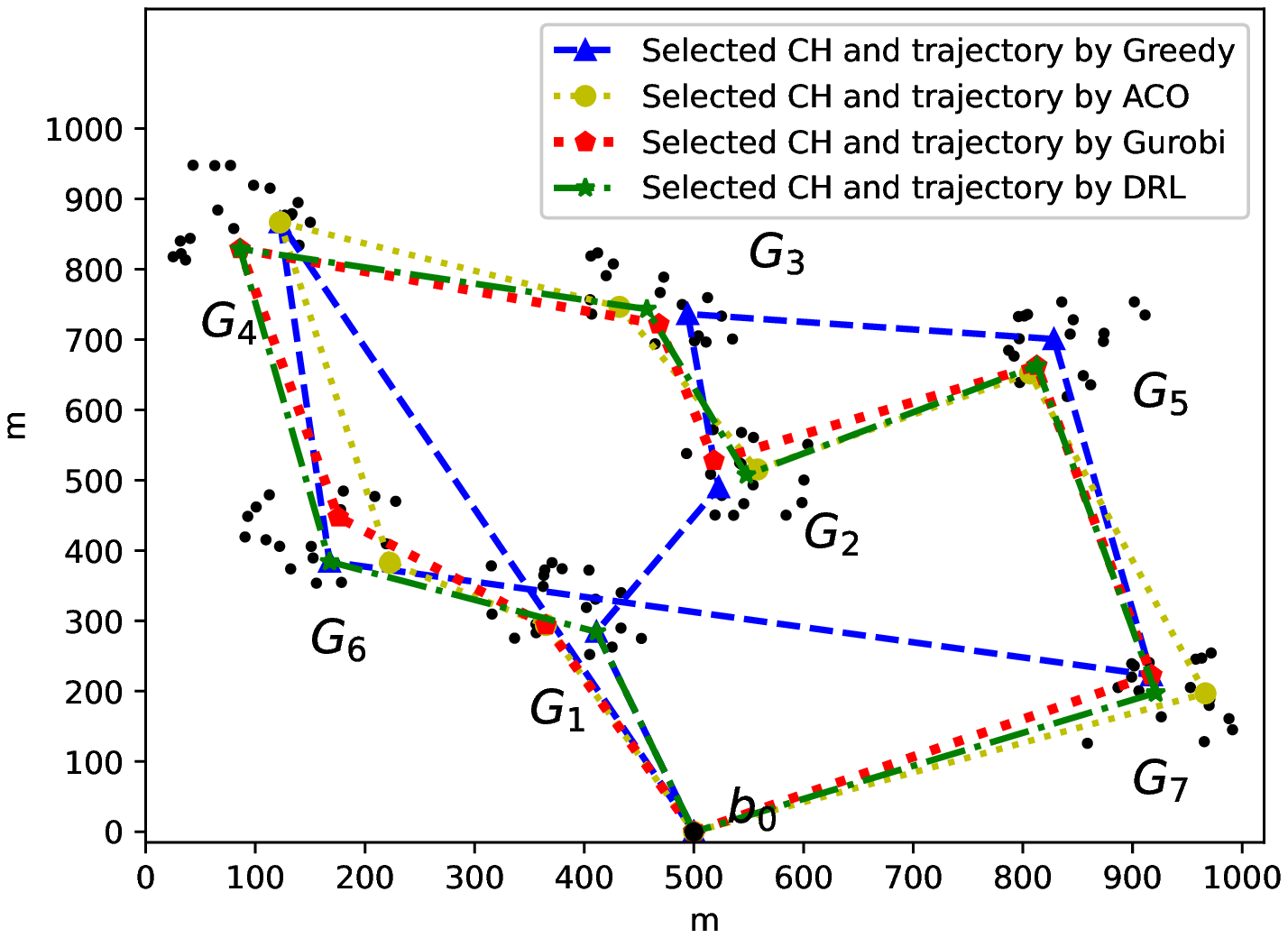}}\hspace{-0.2in}
	\subfigure[]{\includegraphics[width=0.5\textwidth]{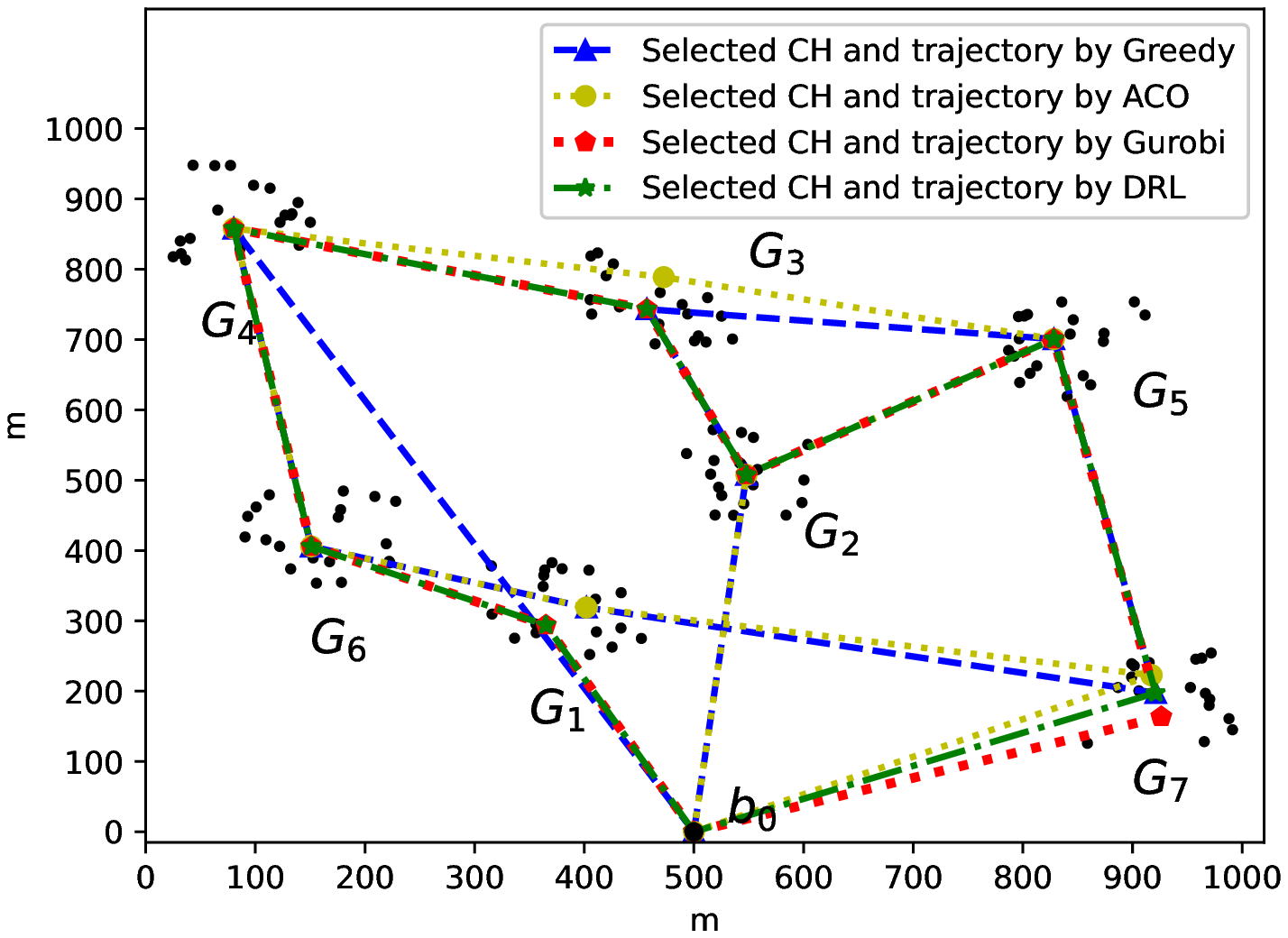}} \\
	\caption{{Trajectory comparison of DRL, greedy algorithm, ACO, and Gurobi for different values of $\omega$. (a) $\omega = 0.1$. (b) $\omega = 0.3$. (c) $\omega = 0.5$. (d) $\omega = 0.8$.}}
	\label{seven clusters trajectory comparison}
	\vspace{0in}
\end{figure*}

For the results in Fig. \ref{trajectory comparison} (d), the value of $\omega$ is set to 0.9.  The greedy algorithm determines CHs and the trajectory according to its local ``greedy'' strategy, {and it finally produces the below access order to the four clusters}
\begin{equation*}
    b_0 \rightarrow G_{4} \rightarrow G_{3}  \rightarrow G_{2} \rightarrow G_{1} \rightarrow b_0.
\end{equation*}
The access order to clusters obtained by ACO is
\begin{equation*}
    b_0 \rightarrow G_{1} \rightarrow G_{2}  \rightarrow G_{4} \rightarrow G_{3} \rightarrow b_0
\end{equation*}
which is better than the order obtained with the greedy algorithm, and inferior to the one found by our proposed DRL algorithm. As for the proposed DRL algorithm, the access order to four clusters is found to be
\begin{equation*}
    b_0 \rightarrow G_{1} \rightarrow G_{3}  \rightarrow G_{4} \rightarrow G_{2} \rightarrow b_0
\end{equation*}
which also nearly coincides with the trajectory obtained by Gurobi.
Through the above four cases, it is clear that our DRL algorithm can find optimal or nearly optimal trajectories when compared with the trajectories found by Gurobi, and it also performs much better than the greedy and the ACO algorithms.

Based on the above simulation results, we present a more detailed analysis of our proposed algorithm in Fig. \ref{comparisonratio}. Specifically, this figure plots the ratios of the energy consumptions of our proposed DRL, the greedy and the ACO algorithms to the energy consumptions of Gurobi which are all normalized to one at different values of $\omega$. The results are averaged over 30 test instances. It can be clearly seen that the energy consumption of our proposed DRL algorithm is very close to the optimal value obtained by Gurobi and less than that of the ACO and the greedy algorithms for all different values of $\omega$. As expected, the ACO algorithm outperforms the greedy algorithm in reducing the energy consumption.


\begin{figure}[!t]
	\centering
	\includegraphics[width=1\linewidth]{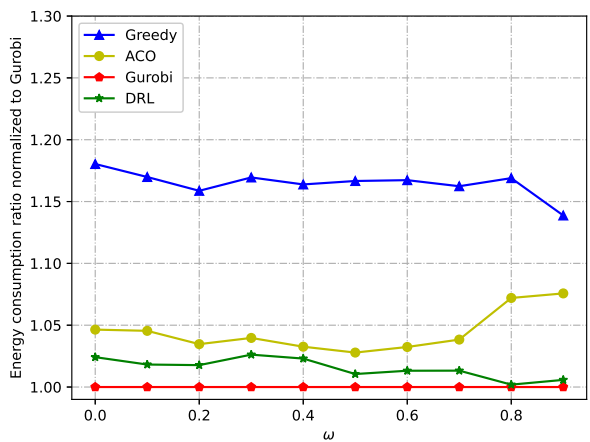}
	\caption{Energy consumption comparison for 7 clusters.}
	\label{comparisonratiocir}
\end{figure}

\subsection{Trajectory and Energy Consumption Comparison on the 7-Clusters IoT Network}

 {In this subsection, we generate a 7-clusters problem instance using the same way as generating the training data, and observe the obtained trajectories on the previously trained 4-clusters model and the three baselines.} As shown in Fig. \ref{seven clusters trajectory comparison},  our proposed DRL algorithm can find the appropriate CH from each cluster even when the value of $\omega$ changes. In addition, it can plan the access trajectory to CHs under different $\omega$, as follows
\begin{equation*}
    b_0 \rightarrow G_{1} \rightarrow G_{6}  \rightarrow G_{4} \rightarrow G_{3} \rightarrow G_{2} \rightarrow G_{5} \rightarrow G_{7} \rightarrow b_0.
\end{equation*}
We can see that the trajectories obtained with the proposed DRL algorithm are very close or fully coincide with the optimal trajectories generated by Gurobi in four cases of Fig.~\ref{seven clusters trajectory comparison}.  However, the greedy algorithm shows the worst trajectory planning ability. In Fig.~\ref{seven clusters trajectory comparison} (a), (b), (c), the UAV's access order to clusters by the greedy algorithm is given by
\begin{equation*}
    b_0 \rightarrow G_{1} \rightarrow G_{2}  \rightarrow G_{3} \rightarrow G_{5} \rightarrow G_{7} \rightarrow G_{6} \rightarrow G_{4} \rightarrow b_0
\end{equation*}
and in Fig.~\ref{seven clusters trajectory comparison} (d), the access order is
\begin{equation*}
    b_0 \rightarrow G_{2} \rightarrow G_{3}  \rightarrow G_{5} \rightarrow G_{7} \rightarrow G_{1} \rightarrow G_{6} \rightarrow G_{4} \rightarrow b_0.
\end{equation*}
As for the ACO algorithm, when compared to the greedy and our proposed DRL algorithms, it can plan a reasonable trajectory to 7 clusters as shown in Fig.~\ref{seven clusters trajectory comparison} (a), (b), (c)
\begin{equation*}
    b_0 \rightarrow G_{1} \rightarrow G_{6}  \rightarrow G_{4} \rightarrow G_{3} \rightarrow G_{2} \rightarrow G_{5} \rightarrow G_{7} \rightarrow b_0
\end{equation*}
but in Fig.~\ref{seven clusters trajectory comparison} (d), its trajectory turns worse, which is given by
\begin{equation*}
    b_0 \rightarrow G_{2} \rightarrow G_{5}  \rightarrow G_{3} \rightarrow G_{4} \rightarrow G_{6} \rightarrow G_{1} \rightarrow G_{7} \rightarrow b_0.
\end{equation*}
The trajectory comparison results show that the model trained by our proposed DRL algorithm has good scalability and generalization abilities to plan the trajectories for new problem instance without retraining the model.

In Fig. \ref{comparisonratiocir}, energy consumption comparison over the averaged results of 30 test instances shows that our proposed DRL algorithm can achieve close-to-minimum energy consumption obtained by Gurobi, and performs better than the greedy and the ACO algorithms. {From the above analysis, one can see that the trained model by a large amount of 4-clusters problem instances can plan a near-optimal trajectory on 7-clusters problem instance, without retraining the model for the test data. This is consistent with the fact that RNNs used in our model have been shown to have very good scalability and generalization\cite{Z. Tu}.}

\subsection{Further Investigation for the Generalization Ability}

\begin{figure}[!t]
		\centering
		\includegraphics[width=1\linewidth]{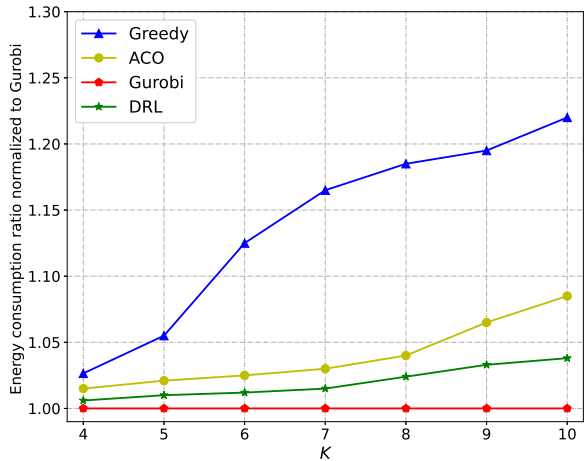}
		\caption{Energy consumption comparison when $K$ varies.}
		\label{ratio}
\end{figure}

\begin{table}[!t]
	\centering
	\caption{{Running time comparison.}}
	\label{timecom}
	 \begin{tabular}{p{0.35cm}<{\centering} | p{0.35cm}<{\centering}|p{1cm}<{\centering}|p{1cm}<{\centering}|p{1cm}<{\centering}|p{1cm}<{\centering}}
		\hline
	   \multicolumn{2}{c|}{} & \multicolumn{4}{c}{{Time}}\\
		\hline
	    \multicolumn{2}{c|}{}  & {Greedy} & {ACO} & {Gurobi} & {DRL}\\
		\hline
	    \multirow{7}*{{$K$}} & {4} & {0.18 s} & {6.92 s} & {1800 s} & {0.36 s}\\
		\cline{2-6}
		& {5} & {0.19 s} & {10.15 s} & {2711 s}  &  {0.37 s}\\
		\cline{2-6}
		& {6} & {0.22 s} & {14.33 s} & {3722 s} &  {0.39 s}\\
		\cline{2-6}
		& {7} & {0.25 s} & {19.23 s} & {5405 s}  & {0.39 s} \\
	    \cline{2-6}
	    & {8} & {0.29 s} & {25.82 s} & {6908 s} & {0.41 s} \\
	    \cline{2-6}
	    & {9} & {0.32 s} & {31.21 s} & {9701 s} & {0.41 s} \\
	    \cline{2-6}
	    & {10} & {0.33 s} & {38.14 s} & {12500 s} & {0.42 s}\\
	    \hline
		\end{tabular}
\end{table}

The scalability and generalization abilities of the trained model is further studied when $K$ varies, and the results are shown in Fig.~\ref{ratio} for $\omega=0.5$. Since Gurobi is the exact solver, it always obtains the optimal solutions at different values of $K$. It is clear that the performance gap between the other three algorithms and Gurobi gradually increases as the value of $K$ increases. However, our proposed DRL algorithm clearly exhibits a superior performance than the greedy and the ACO algorithms in terms of saving the energy consumption. Table \ref{timecom} compares the running times of different algorithms. As the number of clusters increases, the computation times of all four algorithms increase. Although Gurobi obtains the best performance in reducing the energy consumption according to the previous simulation results, it takes the most computational time to deliver the optimal results. The computation time of our DRL algorithm is slightly higher than that of the greedy algorithm, but significantly less than that of the ACO algorithm and Gurobi.

\section{Conclusion}\label{SecVII}

In this paper, we have investigated the problem of jointly designing the UAV's trajectory and selecting CHs for an IoT network to minimize the total energy consumption in the UAV-IoT system. Inspired by the promising development of DRL, we propose a novel DRL-based method to solve this problem. In our proposed method, DRL with a Seq2Seq neural network is used to learn the policy of the trajectory planning with the aim of minimizing the total weighted energy consumption of the UAV-IoT system. Extensive simulation results demonstrated that our proposed method outperforms {the ACO and} greedy algorithms in planning the UAV's trajectory and achieves nearly optimal results when compared to the results obtained by the Gurobi optimizer. In addition, {our proposed DRL algorithm has excellent abilities of generalization, scalability, and automation} to solve different problem instances with different numbers of clusters, without retraining the model for new problems. Considering computation times and the energy consumption results, our proposed method offers an appealing balance between performance and complexity.

\section*{Acknowledgement}

This work was supported by an NSERC/Cisco Industrial Research Chair in Low-Power Wireless Access for Sensor Networks.

\balance

\end{document}